\documentclass{PoS}

\title{Experimental overview of spectroscopy from heavy hadron decays}

\ShortTitle{spectroscopy from heavy hadron decays}

\author{\speaker{Ignacio Bediaga}\thanks{this work of was supported by FAPERG and CNPq Grant No. 157857/2015-
8A.}\\
        CBPF - Centro Brasileiro de Pesquisas F\'isica\\
        E-mail: \email{bediaga@cern.ch}}

\author{Patricia C. Magalh\~aes\\
      CBPF - Centro Brasileiro de Pesquisas F\'isica\\
       }

\abstract{ We present and discuss some experimental approaches involving spectroscopy from heavy meson decays going to light mesons. In particular we emphasis the scalar resonance $f_0(980)$, which have different well determined parameter of mass and width obtained from $J/\psi$ and D  decays. We will also show how the CP violation observed in charmless three body decay can be used as tool to understand hadron hadron interaction at low mass region.}

\FullConference{XVII International Conference on Hadron Spectroscopy and Structure - Hadron2017\\
		25-29 September, 2017\\
		University of Salamanca, Salamanca, Spain}

\begin{document}

\section{Introduction}
Heavy hadron weak decays, mainly non-leptonic ones,  have been used extensively to study light meson spectroscopy in a complementary way to the usual hadron hadron interactions. These decays have a defined initial spin parity quantum numbers that allowed a clear signature to determine the spin  of the intermediary state resonances. Each two body resonance has  a particular distribution in the Dalitz plot associated  to their particular angular momentum quantum number. Beside that, the interference between well known resonances and other amplitude in a particular decay allows the observation of new resonances states, even  if  they  give a small contributions to the total amplitude, due the interference pattern between them. 
  
  However there are  many issues related to a multi-body final state interactions (FSI)   description of heavy hadron decays that are present on the experimental amplitude analysis on these processes: the quasi-to-body approach of the FSI which neglect bachelor particle participation, the Breit-Wigner parameters for mass and width, the unknown non-resonant amplitudes, and the isobar coefficient for magnitude and phase of each individual amplitude, which are constant along the phase space.

 Beside hadron spectroscopy, more recently heavy meson three-body decays has been used as a tool to  search and study CP asymmetry \cite{Bigi1,Bigi2,LHCbcharm,LHCb2014,LHCbppk}. In charm meson decays the observation of CP asymmetry would be a hint of new physics, whereas in charged charmless B  decays observation indicate a expressive CP asymmetry located in some regions of the Dalitz plot. Moreover, the study of the CP asymmetry distribution in these  three-body phase space have been used in one side to understand the mechanism of the CP asymmetry production and on other side to understand  the dynamics of hadron  interactions inside a three-body environment \cite{ LHCb2014,LHCbppk,Tobias1,Tobias2,Tobias3}.  

   In this paper is organized to present in section two an overview of the experimental features and problems related to light  resonance spectroscopy  from charm meson decays, focusing on the scalar  $f_0(980)$ scalar resonance. In section three we present the perspective and inquiring of charmless B meson decay experiments to study light hadron spectroscopy through the large CP asymmetry distribution observed in these decays. In section four we present our final considerations.


\section{Overview on the experimental  light  resonances spectroscopy from charm meson decays.}

The scalar $f_0(980)$ resonance is one of the most universal particle present in these decays, it has a clear signature in all experimental  decays involving two pions or two kaons in the final state. We present an overview of experimental results  involving  two different initial states: one from charmonium $J/\psi$ decays, the other from  weak decays of D and B mesons.

\subsection{The $f_0(980)$ mass and width parameters from $J/\psi$ decays.}

The MarkII\cite{MarkII} collaboration made an  observation of the $f_0(980)$ parameters from a inclusive $J/\psi \pi^-\pi^+ X$  decays. They used a Flatté-Breit-Wigner amplitude distribution to fit the $\pi\pi$ mass spectrum: 
\begin{eqnarray}
BW_{f_0(980)} &=& \frac{1}{m^2_{\pi\pi} - m^2_0 + i\,m_0\,(\Gamma_\pi + \Gamma_K)}\,, \nonumber\\ && \Gamma_\pi = g_\pi \sqrt{m^2_{\pi\pi}/4 - m^2_\pi} \,,\nonumber\\
&& \Gamma_K = \frac{g_K}{2} (\sqrt{m^2_{\pi\pi}/4 - m^2_K} + \sqrt{m^2_{\pi\pi}/4 - m^2_{K_0}} \,.
\label{flatte}
\end{eqnarray}
To perform the fit, they fixed $g_k = 0.2$ and get  $m_0 = 956 \pm 6 $ and  $g_\pi=0.088\pm 0.029$ with the  $3700 \pm 700$  observed events.

    A more complex and complete analysis was performed by BES Collaboration \cite{BES} 
using the exclusives  $J/\psi \to K^-K^+\pi^+\pi^-$ and $J/\psi \to K^-K^+K^-K^+$  channels. They selected events with two kaons in  the $\phi(1020)$ mass region and inspect what remained in  $K^-K^+$ and  $\pi^+\pi^-$ spectrum. With a  partial amplitude  analysis they fitted simultaneously the $J/\psi \to \phi \pi^-\pi^+ $ and $J/\psi \to \phi K^-K^+$ channels.  They constrained the resonance  masses and widths Breit-Wigners parameters  to be the same in both decays for the   resonances amplitudes:  $\sigma, f_2(1270),\, f_0(1370),\, f_0(1500),\, f_2(1525)$ and $ f_0(1710)$.  The $f_0(980)$ was fitted with a  Flatté-Breit-Wigner, Eq.(\ref{flatte}). Other than got the $f_0(980)$ Flatté parameters, they floating also the $f_2(1270), \,f_0(1370),\,  f_2(1525),\, f_0(1790)$ mass and width parameters.

The final adjust for the couple channel partial amplitude analysis results in the following  Flatté-Breit-Wigner  $f_0(980)$ parameters : $m_0 = 965 \pm 10 \pm 15$ MeV and $g_K/g_\pi = 4.21 \pm 0.25 \pm 0.21$. These $f_0(980)$ parameters have been used as an input parameters for other analysis as one can  see on literature.

\begin{figure}[ht]
\begin{center}
\includegraphics[width=.4\columnwidth,angle=0]{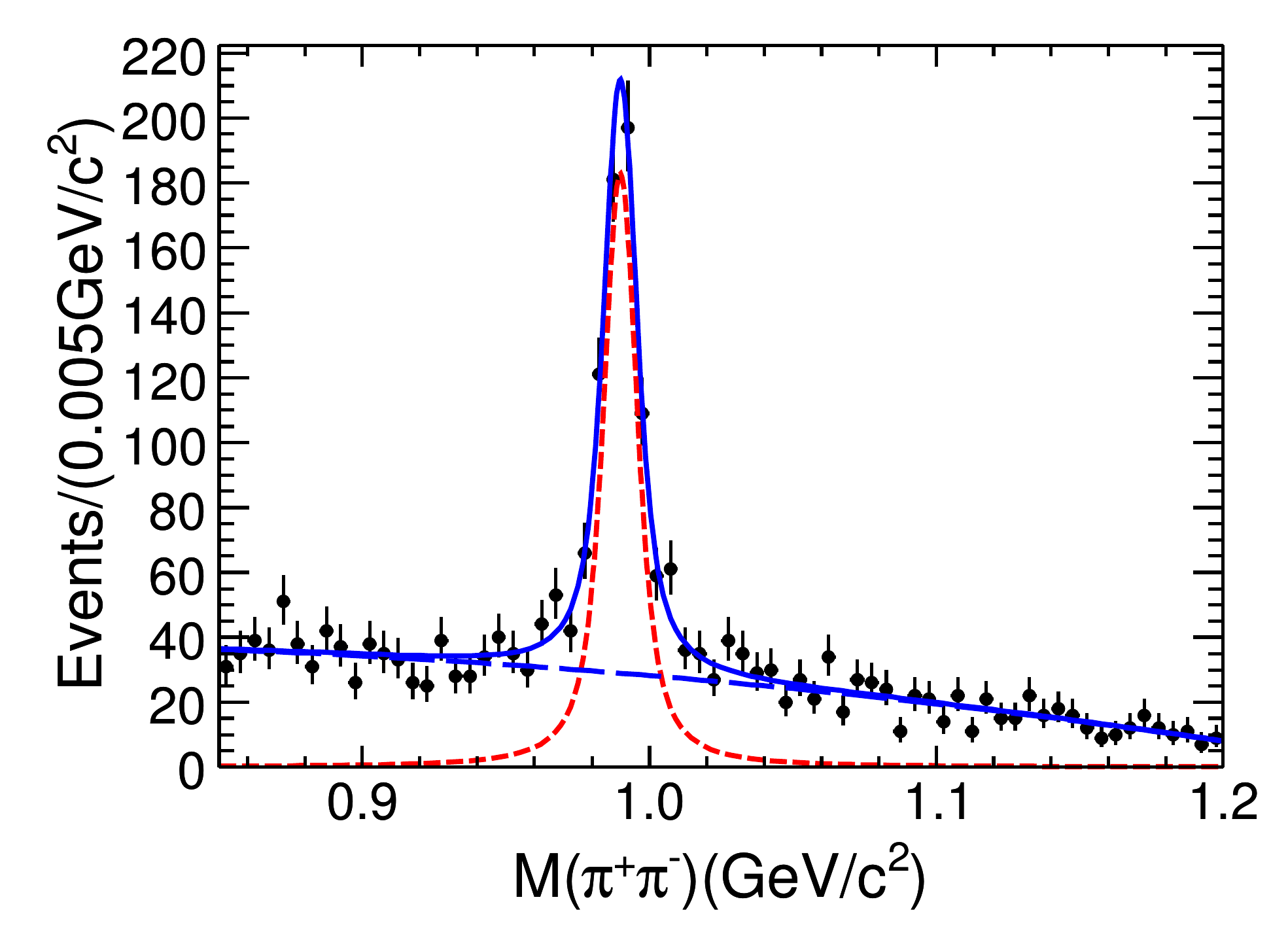}
\caption{ The $\pi^+\pi^-$  invariant mass spectra with $\pi^+\pi^-\pi^0$ in the $\eta(1405)$ mass region. The solid curve is the result of the fit the  dotted curve is
the $f_0(980)$ signal and  dashed one denotes the background polynomial. }
\label{fig:1}
\end{center}
\end{figure}

More recently  BESIII collaboration presented a quite different result for the  $f_0(980)$ parameters from the $J/\psi \to \gamma 3\pi$ decay. In Fig.\ref{fig:1} there is a clear peak in the $3\pi$ invariant mass, associated to the $\eta(1405)$. From  the $ J/\psi \to \gamma \eta(1405)$ events, they look at $\pi^-\pi^+$ invariant mass and observed a clear  $f_0(980)$ signal, much narrow  than the other experimental results for this scalar resonance. The  width parameter measured was $9.5\, \pm\, 1.1$ MeV and mass $989.9\, \pm 0.4$ MeV, these values were obtained with a regular Breit-Wigner. 

The non isospin conservation of this decay, has been attributed as the responsible of this apparent anomaly on the width of   the scalar $f_0(980)$  resonance \cite{Oset}. A similar  decay $J/\psi \to \phi \pi^0\pi^-\pi^+ $, also observed by BESIII collaboration\cite{BESIIIb} with a isospin violation, confirm this narrow $f_0(980)$ resonance.  Direct measurement of the Breit-Wigner  parameters, gave  $m_0 = 989.4  \pm 1.3$ and $\Gamma  = 15.3  \pm 4.7$MeV, compatible with the  previous values from $\eta(1405) \to f_0(980) \pi^0$.

\subsection{The $f_0(980)$ mass and width parameters from charm meson weak decays}

Several three body D weak decays were used  to study the $f_0(980)$ parameters. The pioneer in these studies was the E791 collaboration in  $D_s^+ \to \pi^+\pi^-\pi^+$ decay\cite{E791Ds}. To performed this study they introduced in the Isobaric amplitude Model the possibility to float mass and width of an resonance amplitude. The  $f_0(980)\pi^+$ amplitude has approximately  half of the $D_s^+ \to \pi^+\pi^-\pi^+$  events  with a clear signature in the middle of the Dalitz plot.
This allows E791 even with low statistic, to perform a precise estimative  of $f_0(980)$ parameters using the Flatté-Breit-Wigner parametrization: $m_0 = 977 \pm 3\pm  0.01$ MeV,  $g_\pi= 0.08\pm  0.01\pm 0.01$ and $g_K=0.02 \pm  0.04\pm 0.03$. One can note that the $g_K$ value is compatible with zero  in a clear  opposition to the value obtained by BES from $J\psi$ decays\cite{BES}.  By imposing  $g_K =0$ in their analyses, E791  collaboration\cite{E791Ds} found the  width parameter to be $\Gamma= 44 \pm 3 \pm 2$ MeV.

    This result together with other involving the scalar particles sigma and the kappa, observed respectively in the $D^\pm \to \pi^\pm \pi^+ \pi^-$\cite{sigma} and $D^\pm \to K^\pm \pi^+ \pi^-$\cite{kappa} by the same E791 experiment, open an important and long  discussion related to  the rule  of the  bachelor particle in the three body FSI. In short, the common believe on hadronic three-body decays was that the bachelor particle should be a simple spectator to the other  two hadron interactions and consequently the final amplitude must follow the Watson theorem\cite{watson}  and the phase variation, along the invariant mass of the pair in charm three body phase space, should  be the same as the one observed in elastic scattering experiments. However, as we show bellow, this do not correspond to was observed.

\begin{figure}[ht]
\begin{center}
\includegraphics[width=.6\columnwidth,angle=0]{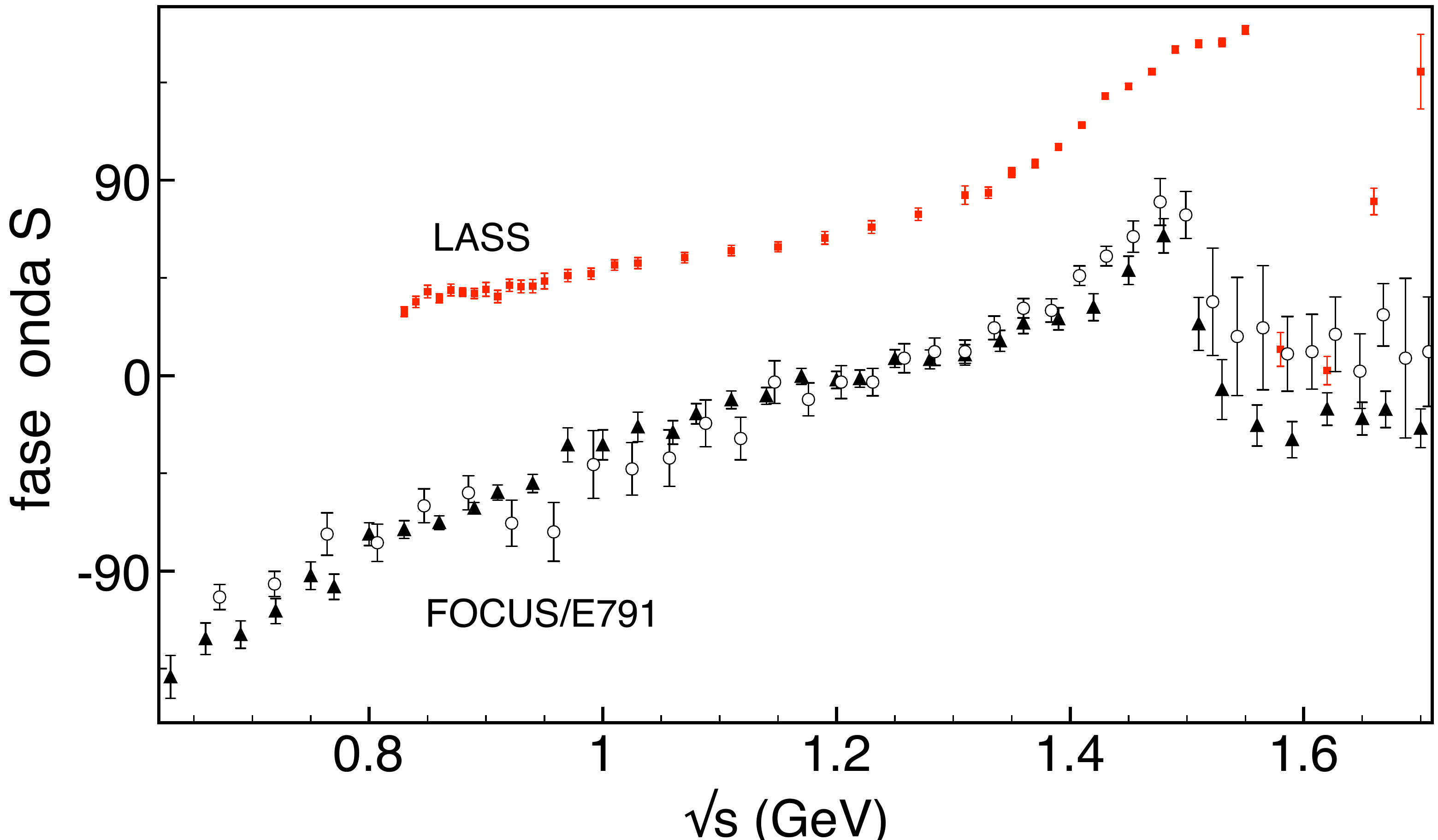}
\caption{ S-wave phase of the $K^-\pi^+$ pair: obtained by FOCUS\cite{FOCUS}(triangle) and E791\cite{kappa} (circle) collaborations in $D^+ \to K^- \pi^+ \pi^+$ decay; and obtained by LASS\cite{Lass} in the free $K^-\pi^+$ scattering. }
\label{fig:watson}
\end{center}
\end{figure}

    In order to test this hypothesis, E791 collaboration redo the   $D^+ \to K^- \pi^+ \pi^+$ analysis, with the same data sample, using a Model Independent Partial Wave Analysis (MIPWA) \cite{E791_pwa}, replacing the scalar resonances kappa and $K_0^*(1430)$ Breit-Wigners, by a continuous S-wave parametriztion  $a(m_{K^-\pi^+})\, e^{- i\phi(m_{K^-\pi^+})}$, dividing the Dalitz plane in rectangular bins. For the spin one and two resonances, it was used the regular Isobar Model, in the same way used on the previous analysis. In Fig.\ref{fig:watson} (left) one can see the results for the phase in the S-wave obtained by this analysis compared with the S-wave phase distribution obtained by scattering amplitude\cite{Lass}.
As we can see from Fig.\ref{fig:watson}  there are an overall difference and also a different dependence in $K^-\pi^+$ invariant mass between the phases, specially at low mass, 
   showing that  Watson theorem\cite{watson} does not work properly  in this decay.
 FOCUS\cite{FOCUS} collaboration performed similar analysis, with high statistic and got  similar behaviour for the S-wave phase motion obtained by E791.

There are some important theoretical works on $D^+ \to K^- \pi^+ \pi^+$ \cite{PRD84, PCMwv,  satoshi, kubisDkpp}. Although based on different frameworks all of them show the importance of three-body FSI to describe the experimental data. In particular, Refs.  \cite{PRD84, PCMwv} have showed that hadron loops introduce new complex structures to the $D^+ \to K^- \pi^+ \pi^+$ amplitude which modify both the  S- and P-wave phase and succeeded in explaining the observed discrepancy presented in Fig.\ref{fig:watson} (left).

\begin{figure}[ht]
\begin{center}
\includegraphics[width=.4\columnwidth,angle=0]{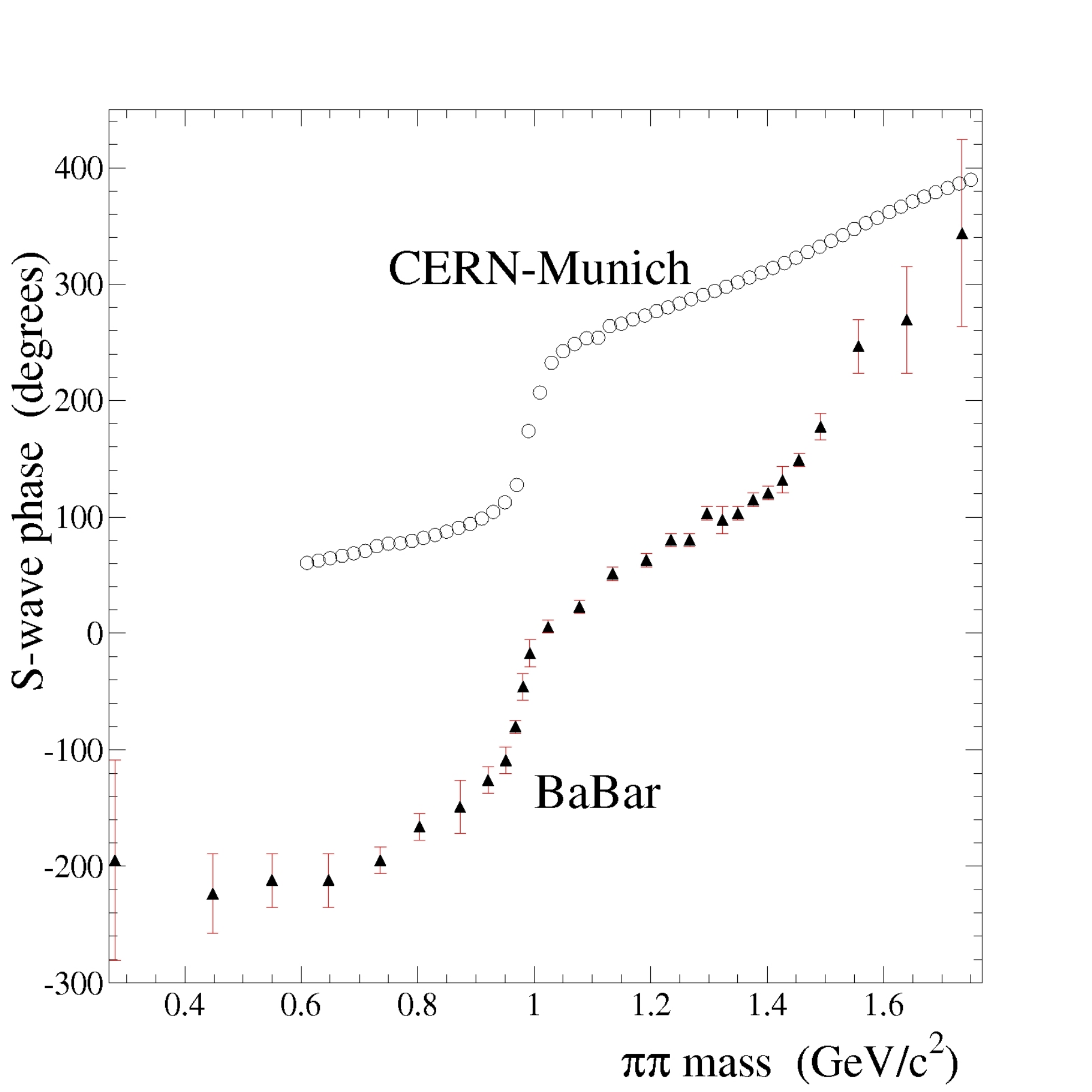}
\caption{ S-wave phase of the $\pi^-\pi^+$ pair: obtained by  Babar \cite{BabarDs}  collaboration in $D_s^+ \to \pi^+\pi^-\pi^+$ decay; and obtained by CERN-Munich\cite{CernMunich} in the free $\pi^-\pi^+$ scattering . }
\label{fig:Dsppp}
\end{center}
\end{figure}

More recently, BaBar collaboration applied the MIPWA technique to the $D_s^+ \to \pi^+\pi^-\pi^+$ decay\cite{BabarDs}. The scalar amplitude extracted from this analysis was  compared with the E791\cite{E791Ds}  with one order of magnitude less events. There is a good  agreement between them in the $f_0(980)$ mass region but not on the entire phase space, being worse to values of $\pi^-\pi^+$  mass square bigger than $1.1$ GeV$^2$. The scalar phase motion obtained with the MIPWA was also compared with the pion pion elastic scattering CERN-Munich  experiment\cite{CernMunich} and it is presented in Figure \ref{fig:Dsppp}\footnote{ want to thanks to Alberto dos Reis for this Figure and  discussions on that matter.}.
Similar to what was observed in the $D^\pm \to K^- \pi^+ \pi^\pm$  decay, Fig. \ref{fig:watson} (right) shows that there is an overall difference of about 200 degrees in $\pi^+ \pi^-$ invariant mass between the pion pion elastic scattering observed by CERN-Munich experiment\cite{CernMunich} and the observed by the BaBar  to  $D_s^+ \to \pi^+\pi^-\pi^+$ decay. Moreover, we can see that the phase motion has a different behavior around the  $f_0(980)$ mass region. Therefore, it is natural that they have a different Breit-Wigner mass and width parameters  as it was pointed out by  $D_s^+ \to \pi^+\pi^-\pi^+$  E791 result.

BaBar collaboration  have other important result in what concerns $f_0(980)$ spectroscopy in charm three-body decay.  They applied the MIPWA to the   $D_s^+ \to K^+ K^-\pi^+$  decay, with high statistic\cite{BabarDsKKp}, to investigate the scalar   $K^+K^-$ invariant mass. The result\cite{BabarDsKKp} shows a smooth phase variation from threshold to $1.15$ GeV, which is complete different from the   $D_s^+ \to \pi^+\pi^-\pi^+$  decay obtained by the same experimental collaboration. Once this analysis was insensible to the  to the coupling of the  with the   $\pi^+\pi^-$ channel, they  used a regular  Breit-Wigner distribution to $f_0(980)$ and obtain from their analysis a  mass parameter of  $m_0= 922 \pm 0.003$ MeV and  the width $\Gamma = 240 \pm 80$ MeV. It is important to point out the these two  results from the same experimental collaboration and same initial state, confirm the importance of the environment to modify the mass and width parameters of a Breit-Wigner like distribution.

In conclusion, for this section, the $f_0(980)$ Breit-Wigner  mass and width  parameters measured  seems to depend strongly on the physics production process involving decays of heavy mesons. Moreover,  analysis performed in charm three-body D decay  showed that FSI are important and Watson theorem can not be apply properly to these three-body decays. The difference between of the phase motion in the elastic scattering to these decays process, have been attributed to the participation of the bachelor particle on the total interaction.  Consequently, if there are  a phase motion different from the scattering, in a  resonance region, is natural that the Breit-Wigner parameters change from one process to the other.

\section{ Perspective of CP violation studies in charmless three body B meson decay}

The huge phase space difference between  charmless three-body B meson decays and charm meson three -body decay rise some questions  yet without a clear responses\cite{ig} Does the nonresonant remain flat? How does it looks like? Does the two body  magnitudes and phases are the same in all phase space? Does the (2+1) approximation works in these decays? Can the entire phase space be described for a single formalism? 

In order to answer these and other questions it would be necessary two crucial elements: a high statistic samples of these decays and a strong collaboration between experimental and theoretical physicists. LHCb collaboration already started both. In fact, $B^\pm \to  K^-K^+K^\pm$ and  $B^\pm \to  K^\pm \pi^+\pi^-$  decays have already a hundred and about two hundred signal events respectively, available in run 1 and probably more than three time this statistics from run 2. LHCb collaboration has been also stimulating  some meetings  between experimental and theoretical physicists in different subjects, but in particular, on the multibody hadron interaction. Some  presentations can be found in the ``annual LHCb series:  Implications of LHCb measurements and future prospects'' and a dedicated workshop on this subjects in Rio de Janeiro\cite{Al_LHCb_ws}.  

There are though some important feature that we have learned from previous Babar, Belle and, more recently, LHCb collaborations. One important thing is that  at low mass the two-body  resonances  are still dominating the total B decay amplitude contributions: $rho(770)$, $K^*(890)$, $\phi(1020)$ and $f_0(980)$,  among others;  like in charm three-body decays. However, differently from  charm decays, the nonresonant contribution in B decays in an important contribution in all channels and  also,  the distribution of this amplitude should not be flat, like it is represented in  charm decays. Although it is not clear how exactly it looks like and what are the mechanism that generate them. 

Charged  charmless B three-meson decays has been presenting a significant inclusive CP violation on data\cite{LHCb2014}. Those decays present a rich CP distribution structure in Dalitz plot, with some  regions  with a high positive CP asymmetry nearby regions also with this kind of asymmetry but with opposite signal as we can see in the figure below. Note that the Fig\ref{fig:LHCb_cpv} is a result of the subtraction of the  negative B meson Dalitz plot for the positive one. The color shows the range of the CP asymmetry. 

\begin{figure}[ht]
\begin{center}
\includegraphics[width=.4\columnwidth,angle=0]{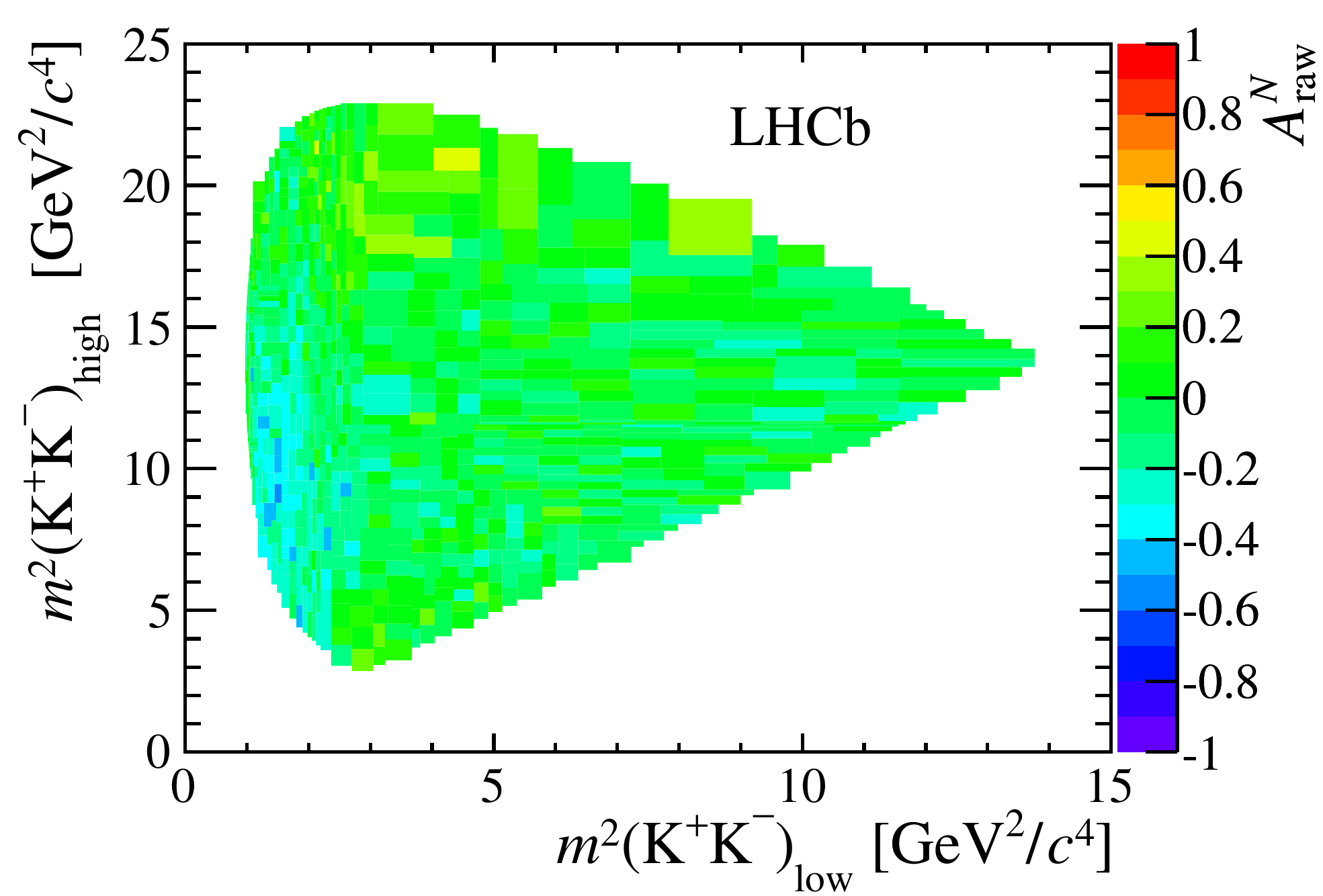}
\includegraphics[width=.4\columnwidth,angle=0]{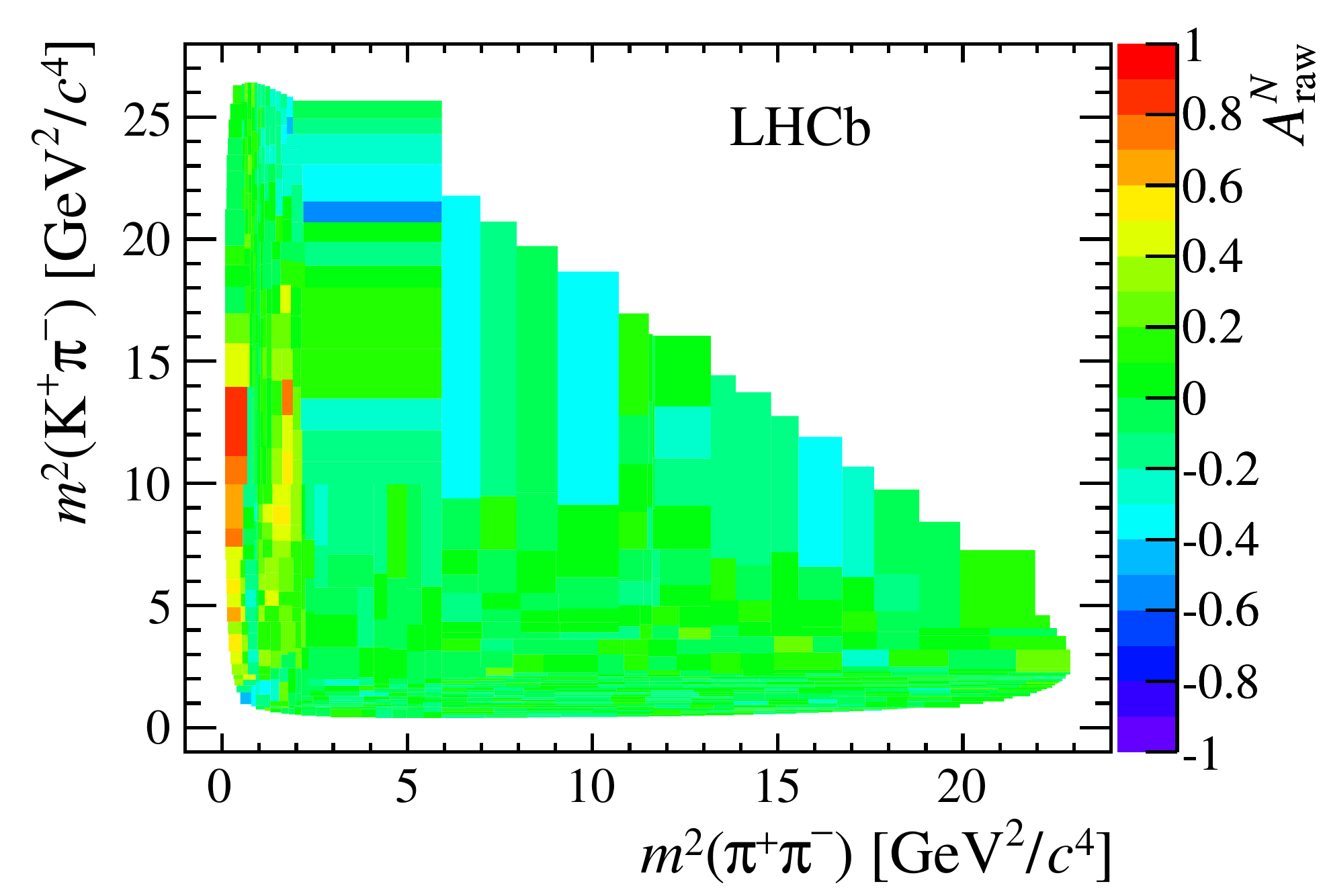}
\includegraphics[width=.4\columnwidth,angle=0]{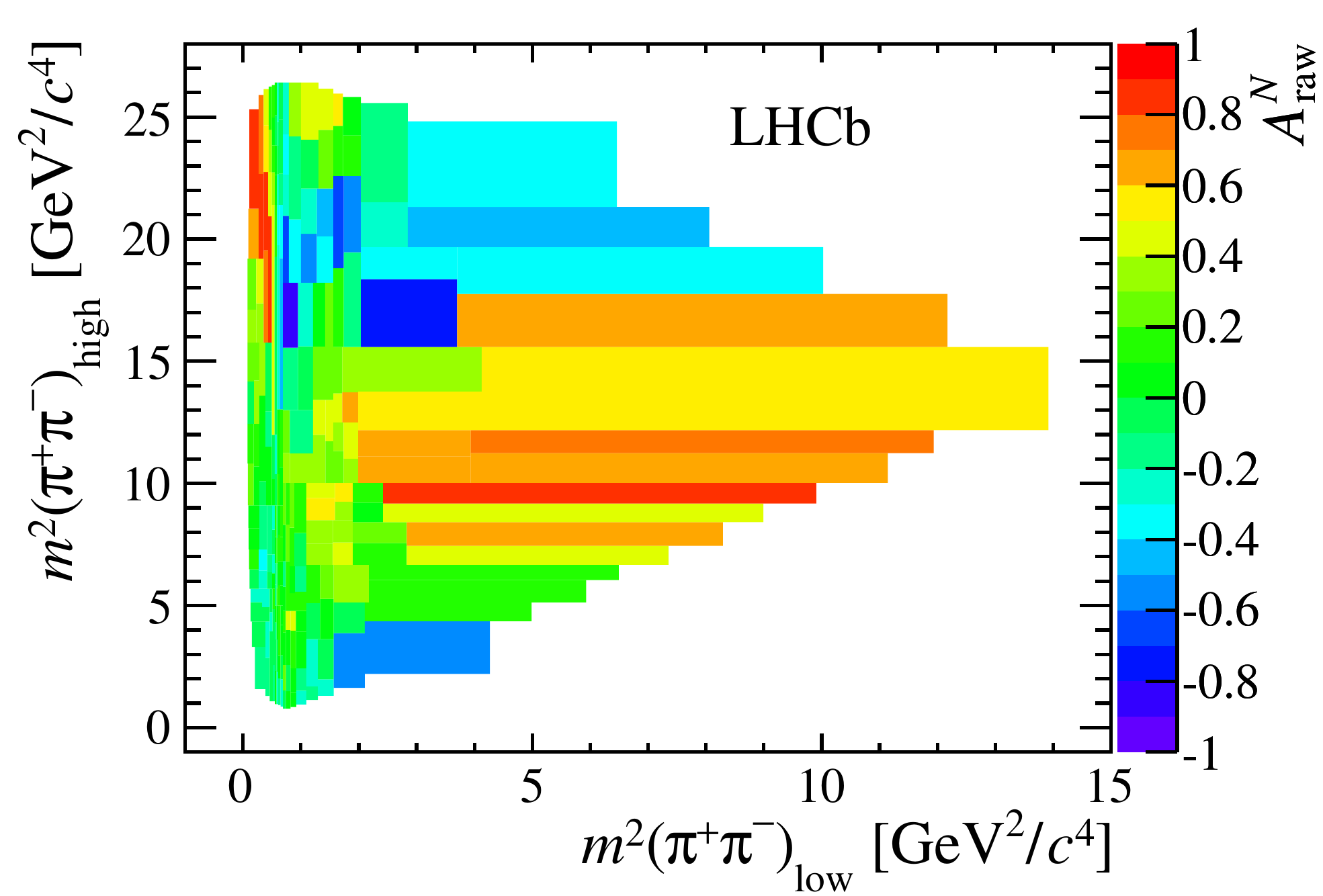}
\includegraphics[width=.4\columnwidth,angle=0]{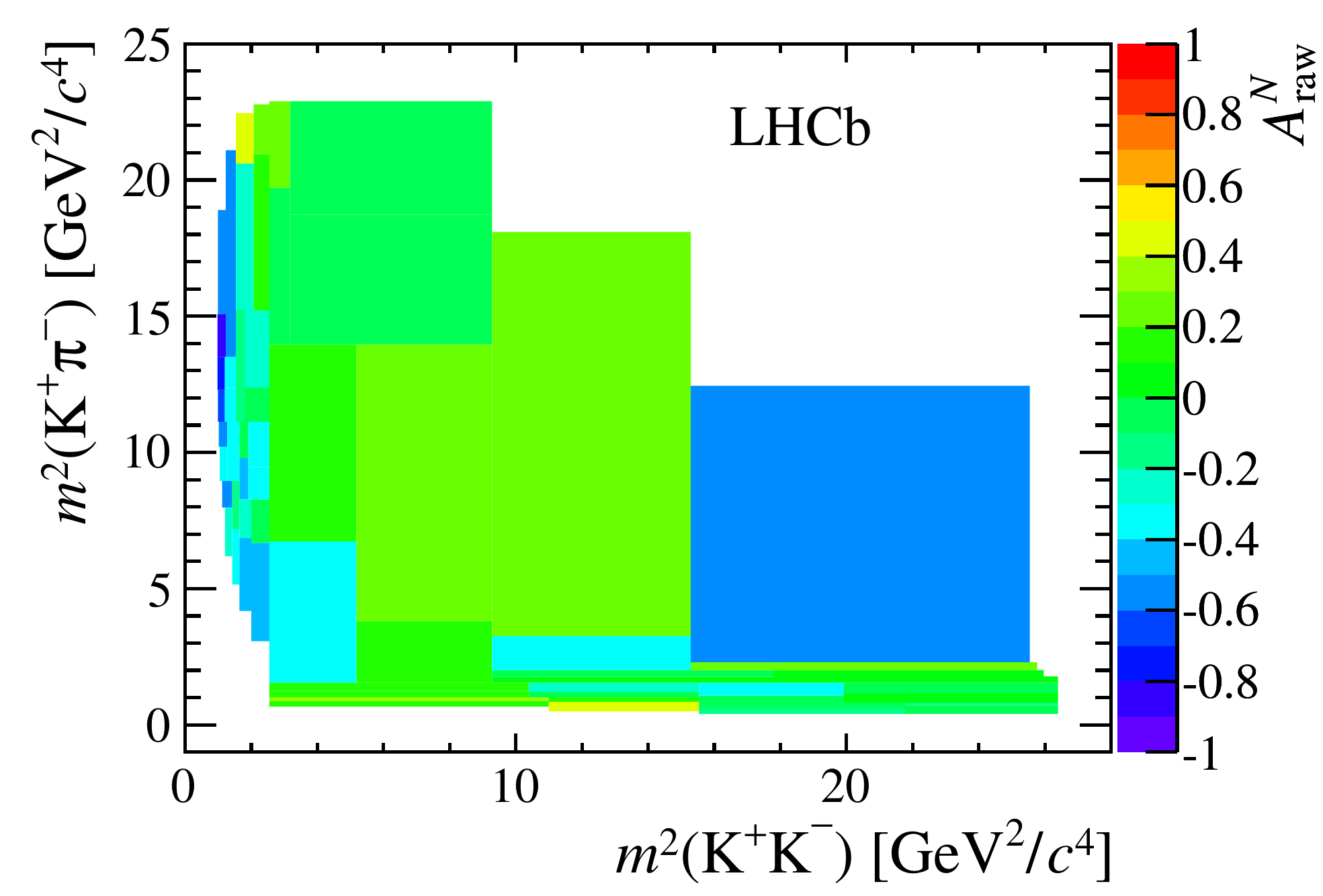}
\caption{ LHCb results on direct CP asymmetry for the channels: $B^\pm \to K^\pm \pi^+ \pi^-$,$B^\pm \to K^\pm K^+ \pi^-$ , $B^\pm \to  \pi^\pm \pi^+ \pi^-$ and $B^\pm \to K^\pm K^+ K^-$ .}
\label{fig:LHCb_cpv}
\end{center}
\end{figure}
%
In order to produce  CP asymmetry the  amplitudes  must have at least two components and these components must have different weak ($\delta$) and strong phases ($\phi$), as given generically by  Eqs.(\ref{eq2.1}) and (\ref{eq2.2}). The  value of this asymmetry  is proportional to the results of Eq.(\ref{eq2.3})
\begin{eqnarray}
A (B\to f)&=& A_1\,e^{i(\delta_1 + \phi_1)} + A_2\,e^{i(\delta_2 + \phi_2)}  \label{eq2.1}
\\
A (\bar{B}\to \bar{f})&=& A_1\,e^{i(\delta_1 - \phi_1)} + A_2\,e^{i(\delta_2 - \phi_2)} \label{eq2.2}
 \\
|A_{B\to f}|^2 - |A_{\bar{B}\to \bar{f}}|^2 &=& -4\,A_1\,A_2\,sin(\delta_1 -\delta_2)\,sin(\phi_1-\phi_2).
\label{eq2.3}
\end{eqnarray}

In general, charmless B decays has a tree diagram with a weak  phase $ \gamma$ from CKM matrix; together with a penguin diagram with a 
quark anti-quart loop. Although the penguin loop is the most common source of strong phase considered in these decays, there could be other sources in the hadronic FSI. Important strong phase difference components can be generated in these decays  though rescattering, Breit-Wigners and the interference between two different partial waves.

The  weak phase signal is defined positive to the particle and negative to the anti-particle, so it is not responsible to the observed CP violation signal changing in the three body B decays. Consequently the  high CP  violation variation observed in Dalitz plane necessarily came from a strong phase variation, allowing a clear signature of this behaviour along the phase space.

It possible to identify two sources of strong phase differences in charmless three body charged B decays in LHCb data\cite{LHCb2014} that can be responsible for the CP asymmetry.  One is related to the $\pi^+\pi^-\to K^+K^-$ rescattering,  and other to the interference between S and P wave at low $\pi^+\pi^-$ invariant mass.  Actually  they showed  a clear correlation  
  between the channels $B^\pm \to K^\pm \pi^+\pi^-$ and  $B^\pm \to K^\pm K^+ K^-$ decays,  observed in the region where 
   $\pi^+\pi^-\to K^+K^-$ has an important contribution in the hadron-hadron scattering amplitude~\cite{Cohen} 
   - i.e. between 1 and 1.6 GeV. The $B^\pm \to K^\pm \pi^+\pi^-$ has a positive CP asymmetry in this region whereas the $B^\pm \to K^\pm K^+ K^-$ has a negative one. A similar correlation in the CP asymmetry, i.e. in the same mass region, was observed between the two channels 
 $B^\pm \to\pi^\pm K^+K^-$ and $B^\pm \to \pi^\pm \pi^+\pi^-$. These results indicate that the re-scattering process  
 $\pi^+\pi^-\to K^+K^-$ is present in these decays~\cite{Tobias1,Tobias2}, carrying the strong phase necessary for CP violation 
 and conserving CPT global symmetry as discussed in Ref.~\cite{Tobias1,Tobias2}. 

Another source of CP violation identified in these data for the low $\pi^+\pi^-$ invariant mass in the  $B^\pm \to K^\pm \pi^+\pi^-$ and $B^\pm \to \pi^\pm \pi^+\pi^-$ decays is related to the interference between S and P wave projections.
\begin{figure}[ht]
\begin{center}
\includegraphics[width=.4\columnwidth,angle=0]{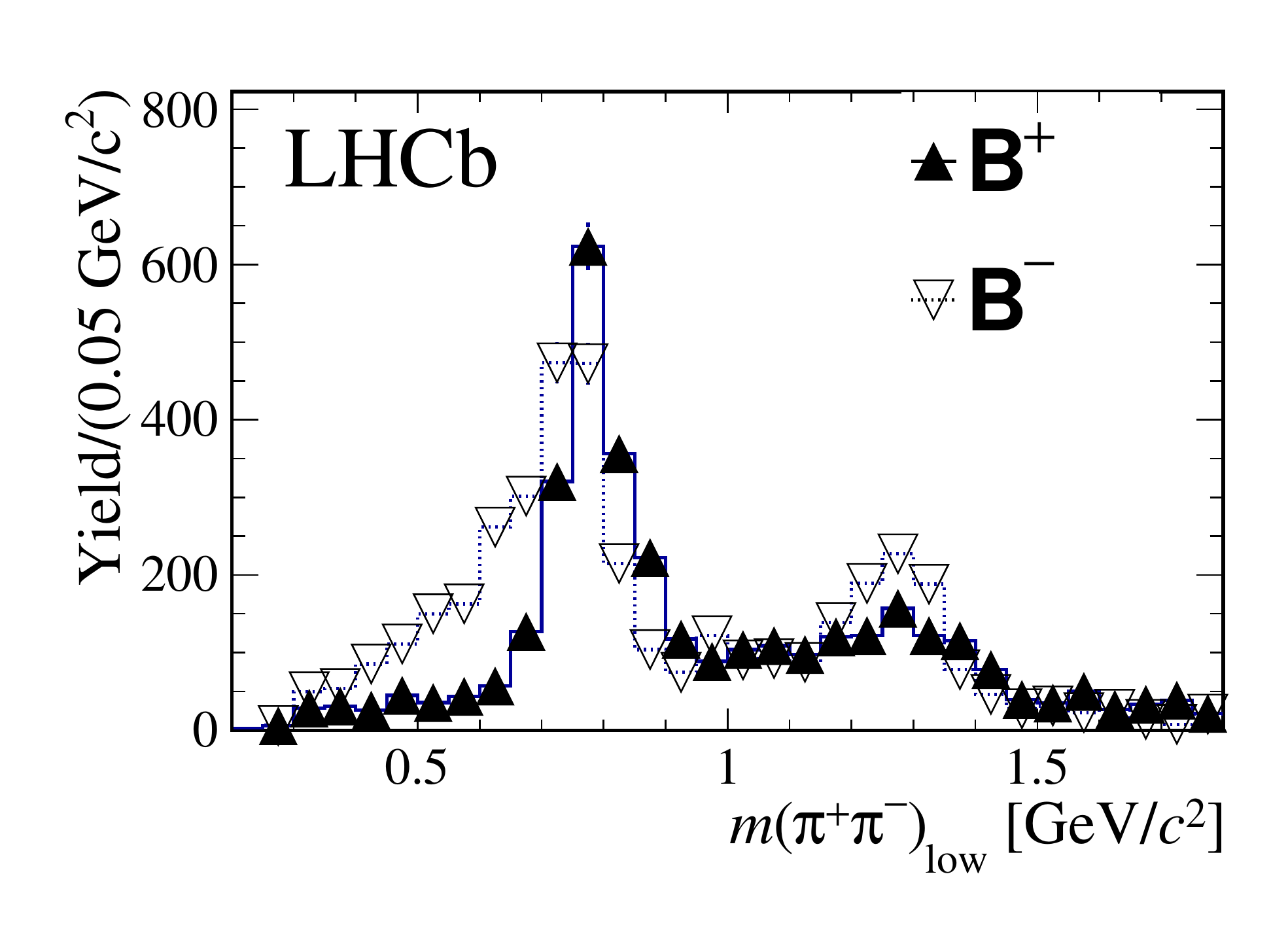}
\includegraphics[width=.4\columnwidth,angle=0]{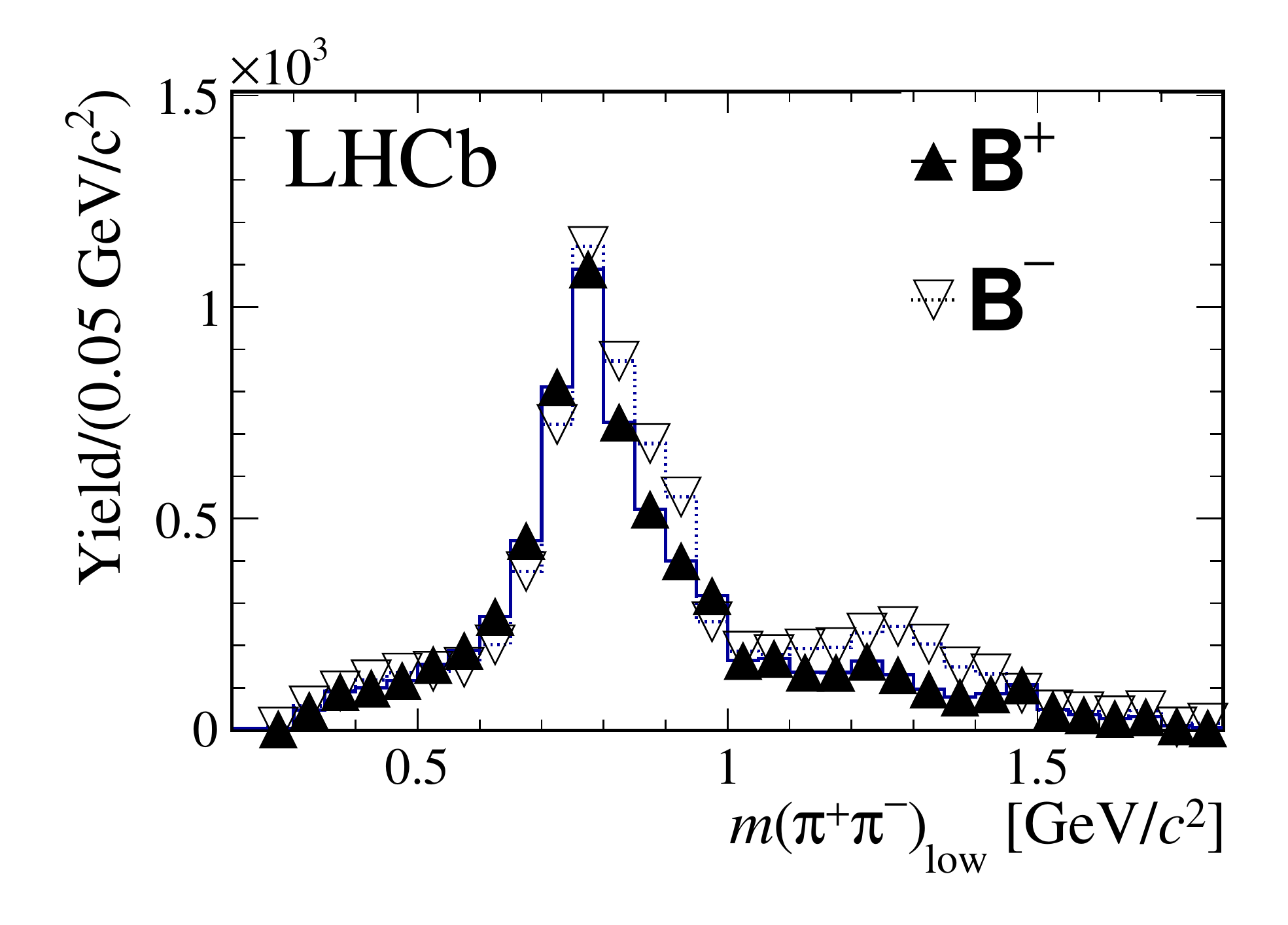}
\includegraphics[width=.4\columnwidth,angle=0]{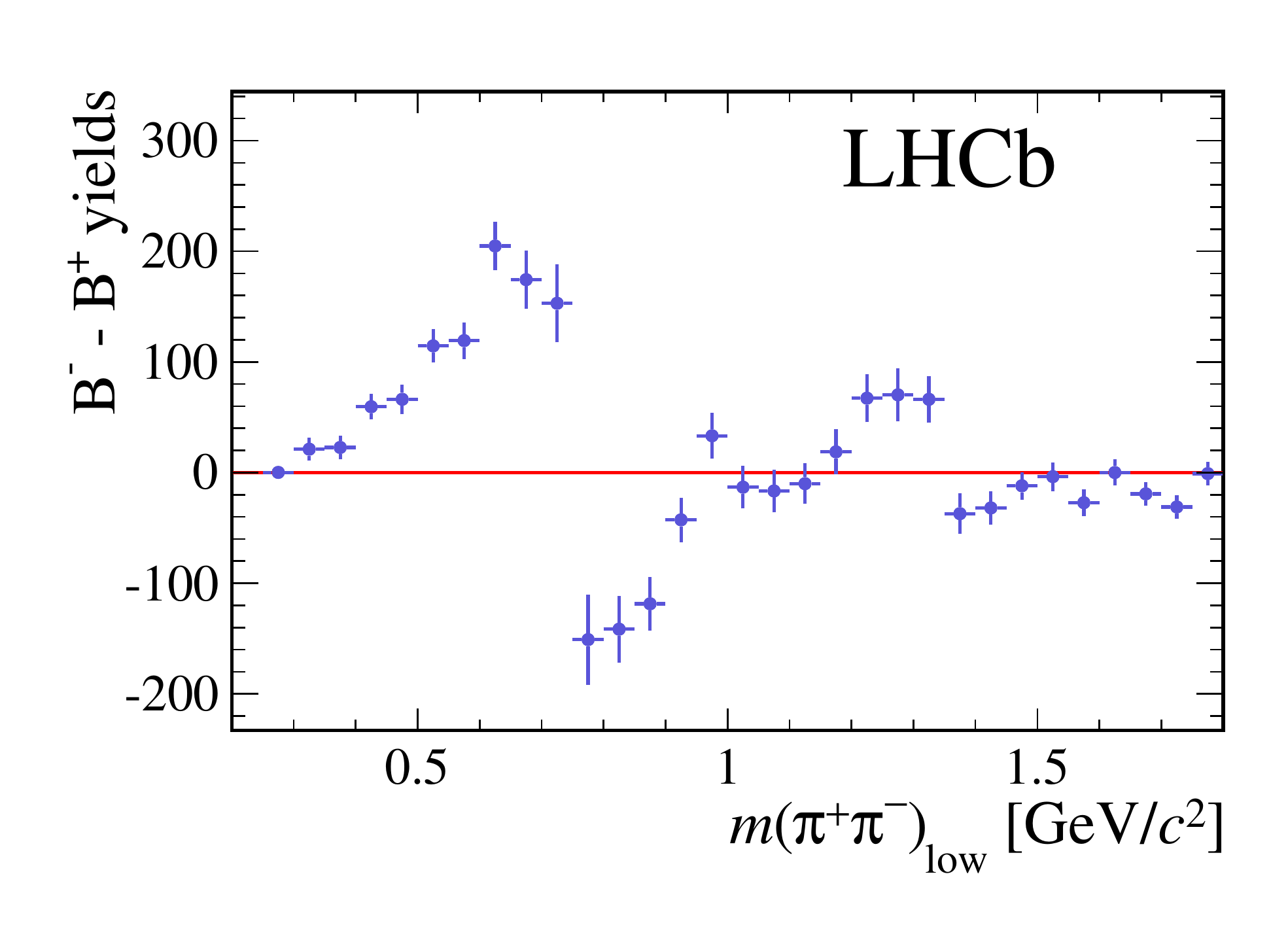}
\includegraphics[width=.4\columnwidth,angle=0]{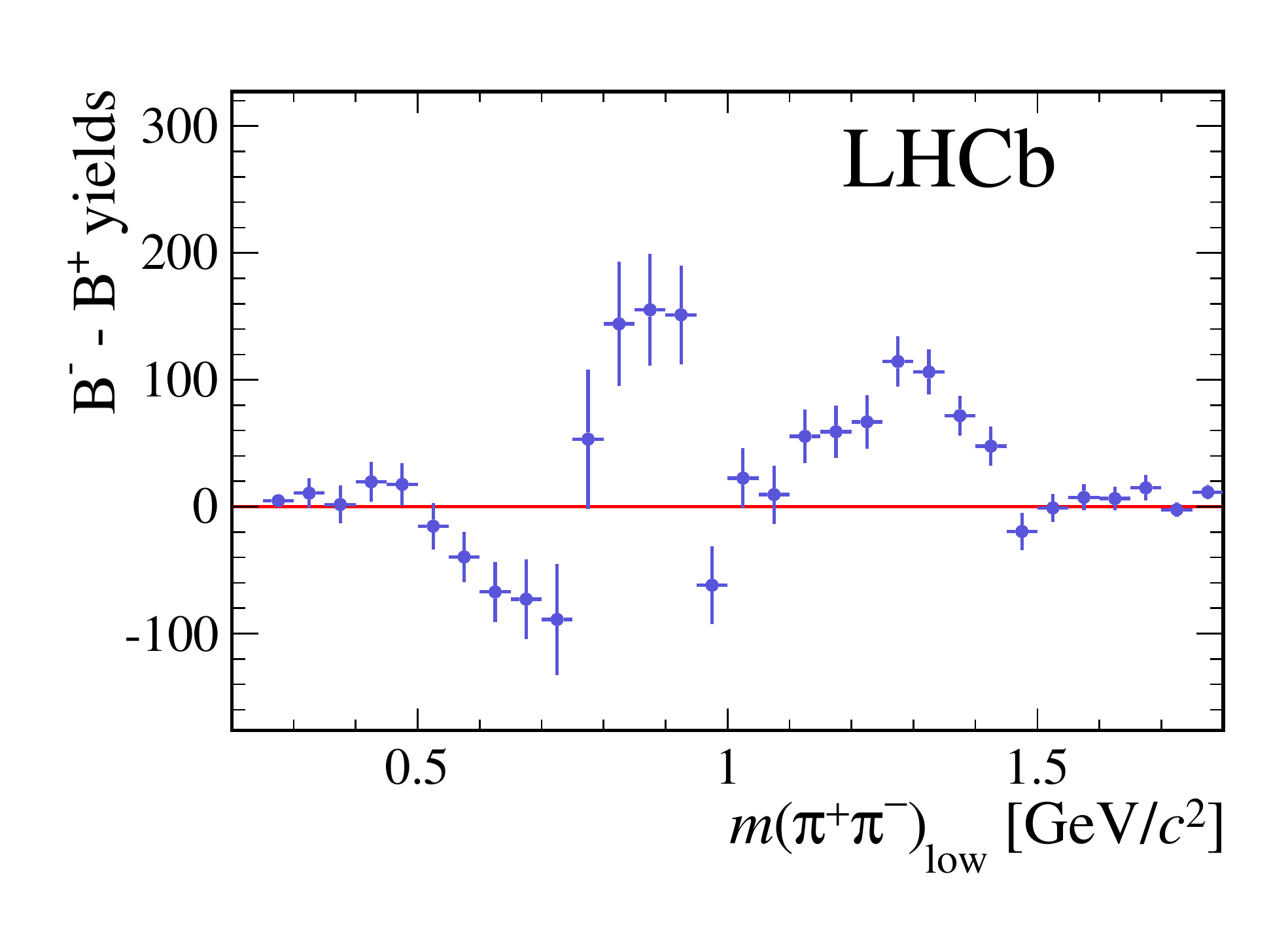}
\caption{Projections in bins of the $m(\pi\pi)_{low}$ variable of (a, b) the number of $B^-$ and $B^+$ signal events and (c, d) their difference for $B^\pm\to\pi^+\pi^-\pi^\pm$ decays.
The plots are restricted to events with (a, c) $\cos\theta<0$  and (b, d) $\cos\theta>0$, with $\cos\theta$ defined in the text. The yields are acceptance-corrected and background-subtracted.
A guide line for zero (horizontal red line) was included on plots (c, d).}
\label{fig:cpvProj}
\end{center}
\end{figure}
Figure \ref{fig:cpvProj}  shows  the  $\pi^+\pi^-$ spectrum  for     $B^- \to \pi^- \pi^+\pi^-$ (open triangle) and $B^+ \to \pi^+ \pi^+\pi^-$ (full triangle) decays,  in two different regions of the Dalitz plot, defined by the vector angular distribution   $cos \theta$. One for $cos \theta>0 $ Fig. \ref{fig:cpvProj}(top-left) and another for $cos \theta<0 $  Fig. \ref{fig:cpvProj} (top-right). On the bottom part of  these Figures we could find the spectrum subtraction. 
One can see  a zero in both distributions at the $\rho^0(770)$ mass distribution position, indicating that the interference term between a S-wave and vector $\rho^0(770)$ is dominated by the real term of the Breit-Wigner. The signal exchange between  Fig.\ref{fig:cpvProj}(top-left) and Fig.\ref{fig:cpvProj}(top-right) shows that this interference term is proportional to the $cos \theta $, which vary between $-1$ to $1$ in the spectrum, as one can expect from a S and P wave interference. A similar process can be seen in $\pi^+\pi^-$ low mass spectrum of the $B^\pm \to K^\pm \pi^+\pi^-$ decay.  

\begin{figure}[ht]
\begin{center}
\includegraphics[width=.4\columnwidth,angle=0]{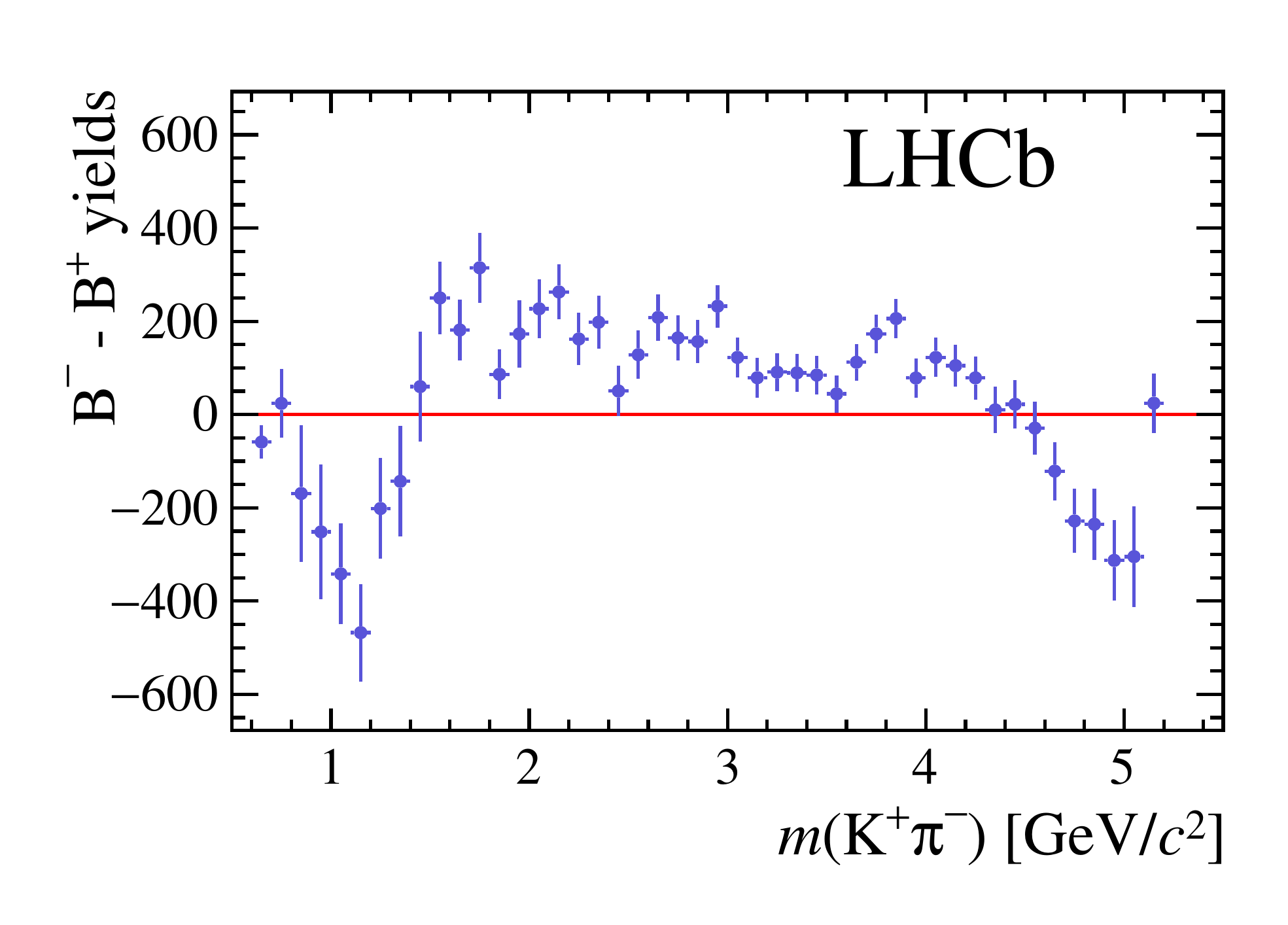}
\includegraphics[width=.4\columnwidth,angle=0]{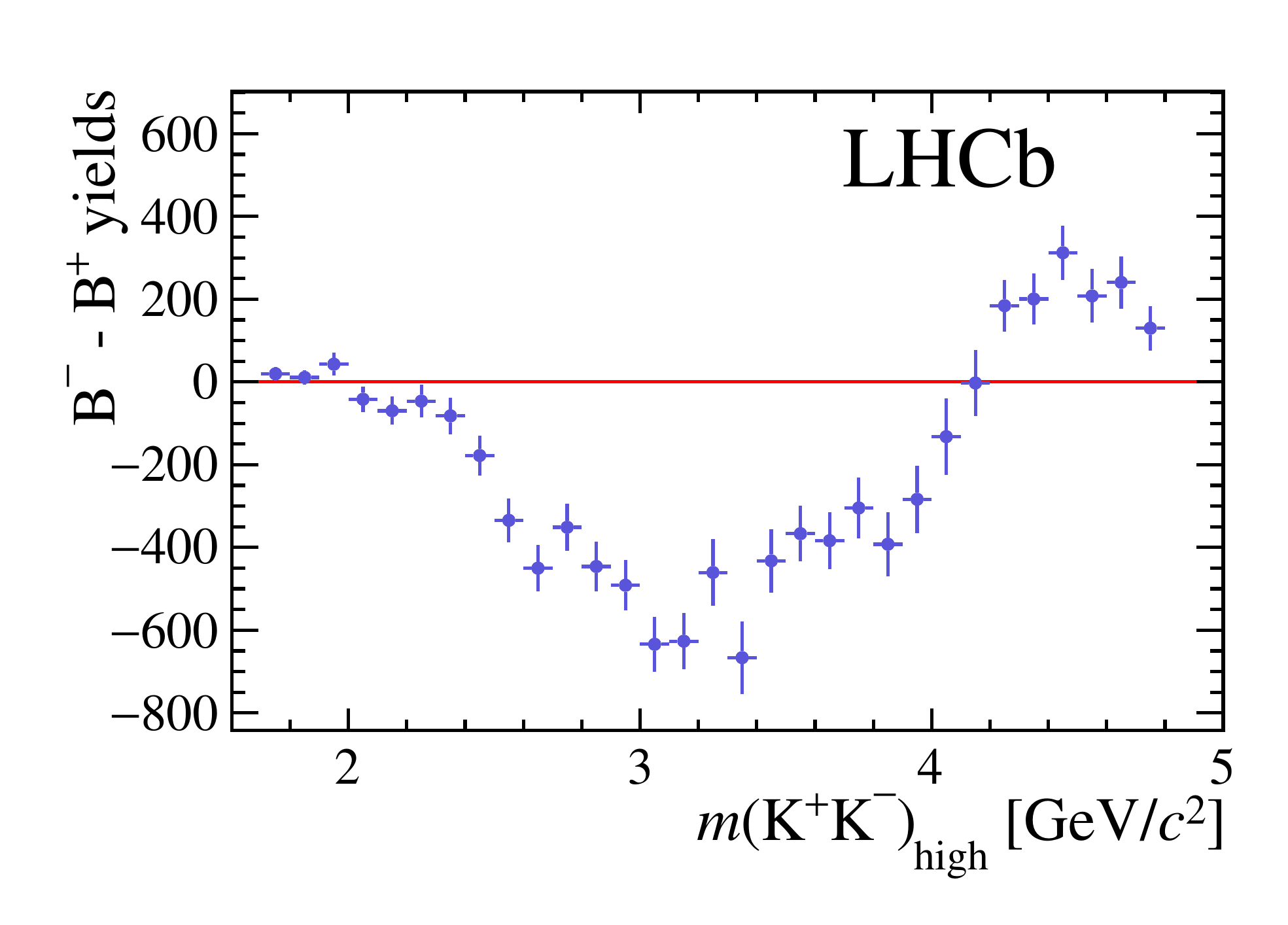}
\includegraphics[width=.4\columnwidth,angle=0]{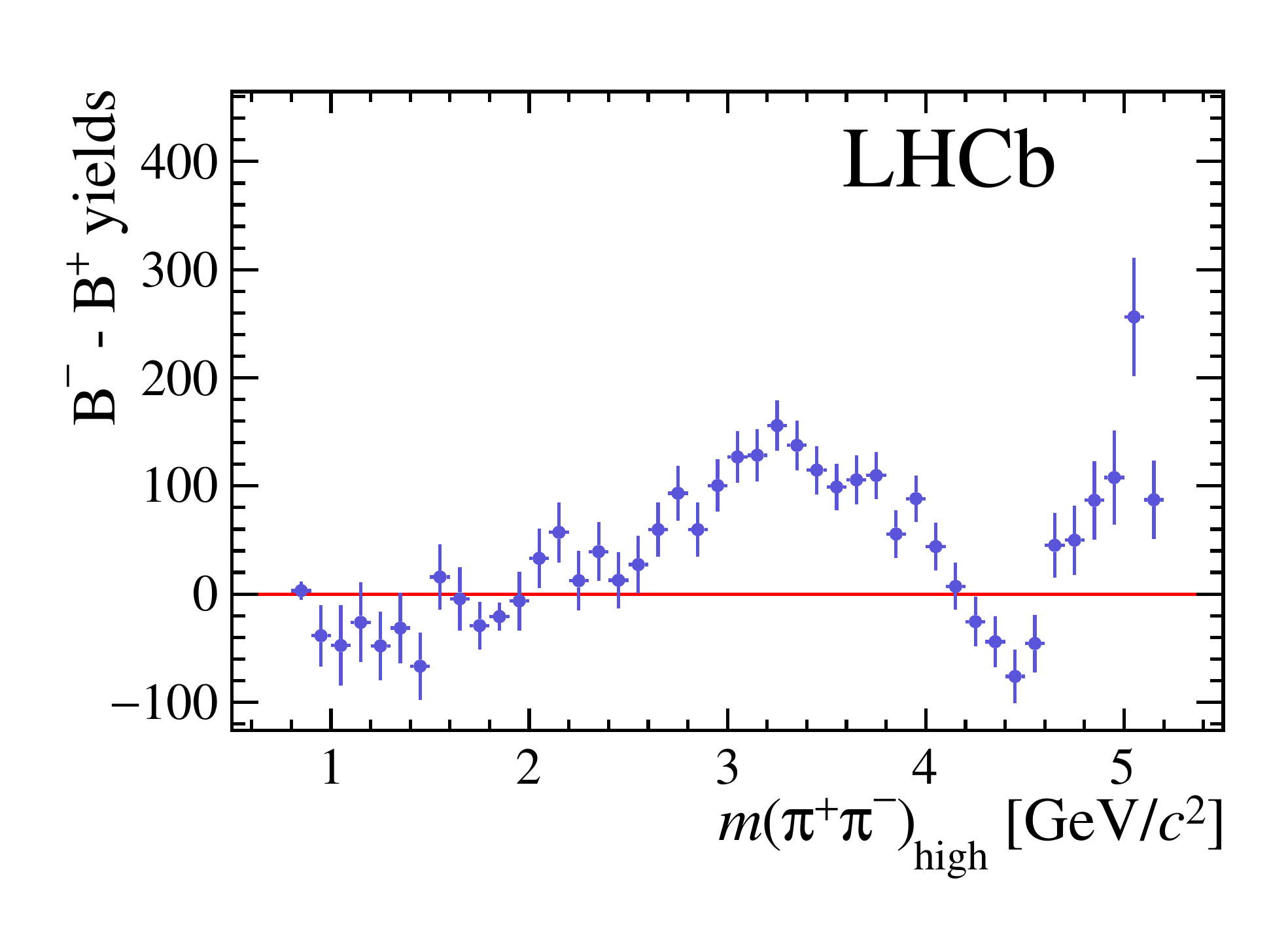}
\includegraphics[width=.4\columnwidth,angle=0]{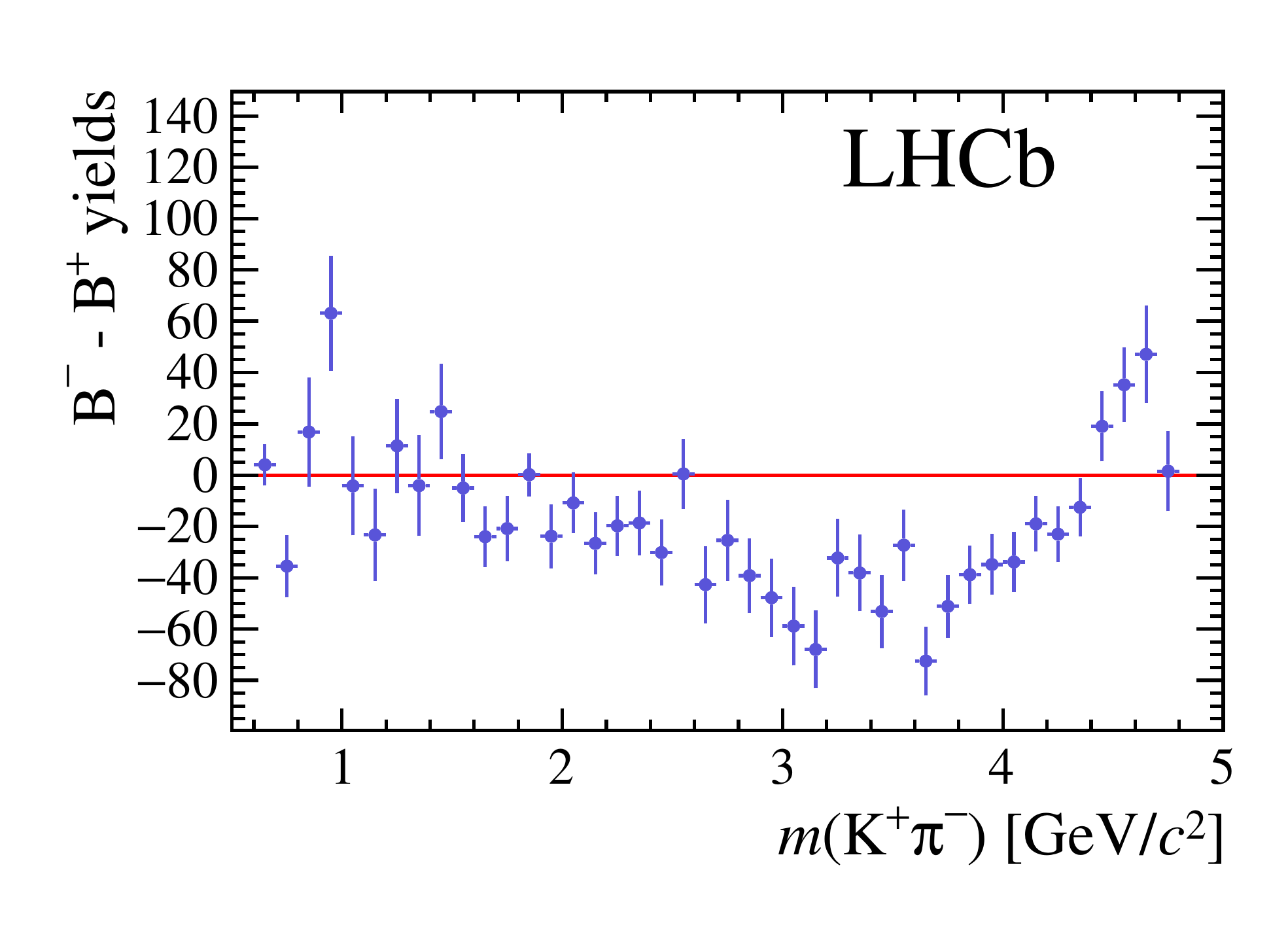}
\caption{ CP asymmetry projection on $m(hh)$ (similar to the previous figure for all phase -space) for the the channels:  $B^\pm \to K^\pm \pi^+\pi^-$ (top-left),   $B^\pm \to K^\pm K^+K^-$ (top- right),  $B^\pm \to \pi^\pm \pi^+\pi^-$ (bottom- left ), $B^\pm \to \pi^\pm K^+K^-$ (bottom- right ).  }
\label{fig:highCPV}
\end{center}
\end{figure}

LHCb data\cite{LHCb2014} also present CP asymmetries placed in high mass regions of the Dalitz plane, near the open channel of charm mesons.
In Fig.~\ref{fig:highCPV} we access the amount of LHCb ~\cite{LHCb_addInfos} events related to CP violation in the four channels $B^\pm \to h^\pm h^+ h^-$ obtained by the subtraction of events for $B^+$ and $B^-$ integrated in $m(hh)$. 
 On can see from Fig~\ref{fig:highCPV} that in all channels the CP asymmetry changes sign crossing zero  at $\approx 4$ GeV. 
  Moreover, it is possible to observes a correlation between the two top and bottom  graphs: they present an opposite direction of CP asymmetry sign change. 
  Inspired by the $\pi\pi \to KK$ rescattering and by the CP sign change near the $D\bar{D}$ open channel,  we investigate in a recent paper\cite{full} the possibility of the double charm rescattering to light mesons to be a source of a new strong phase variation at high mass that could be a new mechanism of CP violation on those decays.

\section{Final remarks}
Three body charm meson decays have been given many important contribution for light meson spectroscopy, mainly to understand scalar resonances. With the order of millions of charm decays events per  channel,  LHCb experiment can go further and look at double Cabibbo decays with detail other than redo interesting decays like  $D^\pm_s \to \pi^\pm \pi^+\pi^-$  to confirm previous results. 

On the other hand, charmless three body B decays already have channels with high statistics, but with many open question raised due to the huge phase space. Thus we have to learn with data and new theoretical inputs. LHCb experiment already have  $ 3ft^{-1}$ in run 2, collected with 13 TeV in the center of mass energy. At this energy, the $ b\bar b $ cross section is about twice the run 1\cite{Bcross}. Therefore, we can expect  soon a LHCb experimental results with around half million events to  $B^\pm \to K^\pm \pi^+\pi^-$ decay. 

With this  statistics and the big amount of CP violation observed in charmless three body B meson decays, we can go deeply to understand many issues involving strong phases along the phase space of these decays.  Also  understand better the necessary tools and theoretical issues to perform a realistic amplitude analysis, necessary to extract physical quantities.

\end{document}